\documentclass{elsart3p}

\usepackage{graphicx}
\usepackage{epsfig}

\usepackage{amssymb}

\begin{document}

\begin{frontmatter}




\title{Electronic structure of the electron-doped cuprate superconductors}


\author{Li Cheng, Huaiming Guo, and Shiping Feng}

\address{Department of Physics, Beijing Normal University, Beijing 100875,
China}

\begin{abstract}
Within the framework of the kinetic energy driven d-wave
superconductivity, the electronic structure of the electron doped
cuprate superconductors is studied. It is shown that although there
is an electron-hole asymmetry in the phase diagram, the electronic
structure of the electron-doped cuprates in the
superconducting-state is similar to that in the hole-doped case.
With increasing the electron doping, the spectral weight in the
$(\pi,0)$ point increases, while the position of the superconducting
quasiparticle peak is shifted towards the Fermi energy. In analogy
to the hole-doped case, the superconducting quasiparticles around
the $(\pi,0)$ point disperse very weakly with momentum.
\end{abstract}

\begin{keyword}
Electron-doped cuprate superconductors\sep Electronic structure\sep
d-wave superconductivity


\PACS 74.25.Jb \ 74.20.Mn \ 74.20.-z

\end{keyword}
\end{frontmatter}

The electron-doped cuprate superconductors are an important
component in the puzzle of the high temperature superconductivity.
The undoped material is a Mott insulator with the antiferromagnetic
(AF) long-range order (AFLRO), then superconductivity emerges when
electrons are doped into this Mott insulator \cite{tokura}. It has
been found that only an approximate symmetry in the phase diagram
exists about the zero doping line between the hole-doped and
electron-doped cuprates \cite{alff}, and the significantly different
behavior of the electron-doped and hole-doped cuprates is observed
\cite{shen}, reflecting the electron-hole asymmetry. In the
electron-doped cuprates, AFLRO survives until superconductivity
appears over a narrow range of the electron doping, around the
optimal doping $\delta\sim 0.15$ \cite{tokura,peng}, where the
commensurate magnetic scattering peak is observed at low and
intermediate energies, with increasing energy this commensurate
magnetic scattering peak is split, and then the incommensurate
magnetic scattering peaks appear at high energy \cite{dai}. By
virtue of systematic studies using the the angle-resolved
photoemission spectroscopy (ARPES) technique, the electronic
structure of the electron-doped cuprates has been well established
\cite{shen,shen1,shen2,shen3,shen4}: (a) in the normal-state, the
charge carriers doped into the parent Mott insulators first enter
into the $[\pi,0]$ (in units of inverse lattice constant) point in
the Brillouin zone, this is different from the hole-doped case,
where the charge carriers are accommodated at the $[\pi/2,\pi/2]$
point; (b) however, in the superconducting (SC)-state, the lowest
energy states are located at the $[\pi,0]$ point, in other words,
the majority contribution in the SC-state for the electron spectrum
comes from the $[\pi,0]$ point. This is the same as in the
hole-doped case; (c) the electron spectrum is characterized by a
sharp SC quasiparticle peak at the $[\pi,0]$ point; and (d) although
the momentum dependence of the SC gap function is obviously deviates
from the monotonic d-wave gap function \cite{matsui}, it is
basically consistent with the d-wave symmetry \cite{shen2,sato} as
in the hole-doped case. Therefore, the investigating similarities
and differences of the electronic structure between the hole-doped
and electron-doped cuprate superconductors would be crucial to
understanding physics of the high temperature superconductivity
\cite{shen,shen1,shen2,shen3,shen4}.

In our earlier work \cite{guo} based on the kinetic energy driven
SC mechanism \cite{feng}, the electronic structure of the
hole-doped cuprates in the SC-state has been discussed, and some
main features of the ARPES experiments on the hole-doped cuprate
superconductors are qualitatively reproduced, including the doping
dependence of the electron spectrum and quasiparticle dispersion
around the $[\pi,0]$ point \cite{shen,campuzano}. In this Letter,
we study the electronic structure of the electron-doped cuprates
in the SC-state along with this line. We show explicitly that
although the electron-hole asymmetry is observed in the phase
diagram \cite{shen,shen1}, the electronic structure of the
electron-doped cuprates in the SC-state is similar to that in the
hole-doped case. With increasing the electron doping, the spectral
weight in the $[\pi,0]$ point increases, while the position of the
SC quasiparticle peak is shifted towards the Fermi energy. In
analogy to the hole-doped case, the SC quasiparticles around the
$[\pi,0]$ point disperse very weakly with momentum.

From the ARPES experiments \cite{shen1,shen5}, it has been shown
that the essential physics of the electron-doped cuprates is
contained in the $t$-$t'$-$J$ model on a square lattice,
\begin{eqnarray}
H&=&t\sum_{i\hat{\eta}\sigma}PC^{\dag}_{i\sigma}
C_{i+\hat{\eta}\sigma}P^{\dag}-t'\sum_{i\hat{\tau}\sigma}
PC^{\dag}_{i\sigma}C_{i+\hat{\tau}\sigma}P^{\dag}\nonumber\\
&-&\mu\sum_{i\sigma}PC^{\dag}_{i\sigma} C_{i\sigma}
P^{\dag}+J\sum_{i\hat{\eta}}{\bf S}_i\cdot{\bf S}_{i+\hat{\eta}},
\end{eqnarray}
with $t<0$, $t'<0$, $\hat{\eta}=\pm\hat{x},\pm\hat{y}$,
$\hat{\tau}=\pm\hat{x} \pm\hat{y}$, $C^{\dagger}_{i\sigma}$
($C_{i\sigma}$) is the electron creation (annihilation) operator,
${\bf S}_{i}= C^{\dagger}_{i}{\vec\sigma}C_{i}/2$ is spin operator
with ${\vec\sigma}=(\sigma_{x},\sigma_{y},\sigma_{z})$ as Pauli
matrices, $\mu$ is the chemical potential, and the projection
operator $P$ removes zero occupancy, i.e.,
$\sum_{\sigma}C^{\dagger}_{i\sigma}C_{i\sigma}\geq 1$. For the
hole-doped case, a charge-spin separation (CSS) fermion-spin
theory has been developed to incorporate the single occupancy
constraint \cite{feng1}. To apply this CSS fermion-spin theory in
the electron-doped cuprates, the $t$-$t'$-$J$ model (1) can be
rewritten in terms of a particle-hole transformation $C_{i\sigma}
\rightarrow f^{\dagger}_{i-\sigma}$ as,
\begin{eqnarray}
H&=&-t\sum_{i\hat{\eta}\sigma}f^{\dag}_{i\sigma}
f_{i+\hat{\eta}\sigma}+t'\sum_{i\hat{\tau}\sigma}
f^{\dag}_{i\sigma}f_{i+\hat{\tau}\sigma}\nonumber\\
&-&\mu\sum_{i\sigma}f^{\dag}_{i\sigma}f_{i\sigma}
+J\sum_{i\hat{\eta}}{\bf S}_i\cdot{\bf S}_{i+\hat{\eta}},
\end{eqnarray}
supplemented by the local constraint $\sum_{\sigma}
f^{\dagger}_{i\sigma}f_{i\sigma}\leq 1$ to remove double
occupancy, where $f^{\dagger}_{i\sigma}$ ($f_{i\sigma}$) is the
hole creation (annihilation) operator, while ${\bf S}_{i}=
f^{\dagger}_{i}{\vec\sigma}f_{i}/2$ is the spin operator in the
hole representation. Now we follow the CSS fermion-spin theory
\cite{feng1}, and decouple the hole operators as, $f_{i\uparrow}=
a^{\dagger}_{i\uparrow}S^{-}_{i}$ and $f_{i\downarrow}=
a^{\dagger}_{i\downarrow}S^{+}_{i}$, with the spinful fermion
operator $a_{i\sigma}=e^{-i\Phi_{i\sigma}}a_{i}$ describes the
charge degree of freedom together with some effects of the spin
configuration rearrangements due to the presence of the doped
electron itself (dressed charge carrier), while the spin operator
$S_{i}$ describes the spin degree of freedom, then the single
occupancy local constraint, $\sum_{\sigma}
f^{\dagger}_{i\sigma}f_{i\sigma} =S^{+}_{i}a_{i\uparrow}
a^{\dagger}_{i\uparrow} S^{-}_{i}+ S^{-}_{i}a_{i\downarrow}
a^{\dagger}_{i\downarrow} S^{+}_{i}= a_{i}a^{\dagger}_{i}
(S^{+}_{i} S^{-}_{i}+S^{-}_{i} S^{+}_{i})=1- a^{\dagger}_{i}
a_{i}\leq 1$, is satisfied in analytical calculations, and the
double dressed fermion occupancy, $a^{\dagger}_{i\sigma}
a^{\dagger}_{i-\sigma}= e^{i\Phi_{i\sigma}} a^{\dagger}_{i}
a^{\dagger}_{i} e^{i\Phi_{i-\sigma}}=0$ and $a_{i\sigma}
a_{i-\sigma}= e^{-i\Phi_{i\sigma}}a_{i}a_{i} e^{-i\Phi_{i-\sigma}}
=0$, are ruled out automatically. It has been shown that these
dressed charge carrier and spin are gauge invariant \cite{feng1},
and in this sense, they are real and can be interpreted as the
physical excitations \cite{feng1,laughlin}. Although in common
sense $a_{i\sigma}$ is not a real spinful fermion, it behaves like
a spinful fermion. In this CSS fermion-spin representation, the
low-energy behavior of the $t$-$t'$-$J$ model (2) can be expressed
as,
\begin{eqnarray}
H&=&-t\sum_{i\hat{\eta}}(a_{i\uparrow}S^{+}_{i}
a^{\dagger}_{i+\hat{\eta}\uparrow}S^{-}_{i+\hat{\eta}}+
a_{i\downarrow}S^{-}_{i}a^{\dagger}_{i+\hat{\eta}\downarrow}
S^{+}_{i+\hat{\eta}})\nonumber\\
&+&t'\sum_{i\hat{\tau}}(a_{i\uparrow}S^{+}_{i}
a^{\dagger}_{i+\hat{\tau}\uparrow}S^{-}_{i+\hat{\tau}}+
a_{i\downarrow}S^{-}_{i}a^{\dagger}_{i+\hat{\tau}\downarrow}
S^{+}_{i+\hat{\tau}}) \nonumber \\
&-&\mu\sum_{i\sigma}a^{\dagger}_{i\sigma}a_{i\sigma}+J_{{\rm eff}}
\sum_{i\hat{\eta}}{\bf S}_{i}\cdot {\bf S}_{i+\hat{\eta}},
\end{eqnarray}
with $J_{\rm {eff}}=(1-\delta)^2J$, and $\delta=\langle
a^{\dag}_{i\sigma}a_{i\sigma}\rangle=\langle a^{\dag}_{i}a_{i}
\rangle$ is the electron doping concentration. As in the
hole-doped case \cite{feng,feng1}, the magnetic energy term in the
$t$-$t'$-$J$ model is only to form an adequate spin configuration
\cite{anderson}, while the kinetic energy term has been
transferred as the interaction between the dressed charge carriers
and spins. For the hole-doped case, we \cite{feng} have shown that
the interaction from the kinetic energy term in the $t$-$J$ type
model is quite strong, and can induce the dressed charge carrier
pairing state by exchanging spin excitations in the higher power
of the doping concentration, then the electron Cooper pairs
originating from the dressed charge carrier pairing state are due
to the charge-spin recombination, and their condensation reveals
the SC ground-state. Moreover, this SC-state is controlled by both
SC gap function and quasiparticle coherence, which leads to that
the maximal SC transition temperature occurs around the optimal
doping, and then decreases in both underdoped and overdoped
regimes. Based on this kinetic energy driven SC mechanism
\cite{feng}, we \cite{ma} have also discussed superconductivity in
the electron-doped cuprates, and the result shows that the maximum
achievable SC transition temperature in the optimal doping in the
electron-doped cuprate superconductors is much lower than that of
the hole-doped case due to the electron-hole asymmetry. Following
their discussions \cite{feng,ma}, we can define the SC order
parameter for the electron Cooper pair in the electron-doped
cuprate superconductors as,
\begin{eqnarray}
\Delta&=&\langle C^{\dag}_{i\uparrow}
C^{\dag}_{i+\hat{\eta}\downarrow}- C^{\dag}_{i\downarrow}
C^{\dag}_{i+\hat{\eta}\uparrow}\rangle\nonumber\\
&=&\langle a_{i\uparrow}
a_{i+\hat{\eta}\downarrow}S^{\dag}_{i}S^{-}_{i+\hat{\eta}}-
a_{i\downarrow}a_{i+\hat{\eta}\uparrow}S^{-}_{i}
S^{+}_{i+\hat{\eta}}\rangle \nonumber\\
&=& -\chi_{1}\Delta_{h},
\end{eqnarray}
with the spin correlation function $\chi_{1}=\langle S^{+}_{i}
S^{-}_{i+\hat{\eta}}\rangle$, and the charge carrier pairing order
parameter $\Delta_{h}=\langle a_{i+\hat{\eta}\downarrow}
a_{i\uparrow}-a_{i+\hat{\eta}\uparrow}a_{i\downarrow}\rangle$,
then the full dressed charge carrier diagonal and off-diagonal
Green's functions of the electron-doped cuprate superconductors
satisfy the self-consistent equations as \cite{feng,ma},
\begin{eqnarray}
g({\bf k},\omega)&=&g^{(0)}({\bf k},\omega)+g^{(0)}({\bf k},
\omega)[\Sigma^{(a)}_{1}({\bf k},\omega)g({\bf k},\omega)
\nonumber\\
&-& \Sigma^{(a)}_{2}(-{\bf k},-\omega)\Im^{\dagger}({\bf k},
\omega)], \\
\Im^{\dagger}({\bf k},\omega)&=&g^{(0)}(-{\bf k},-\omega)
[\Sigma^{(a)}_{1}(-{\bf k},-\omega)\Im^{\dagger}(-{\bf k},
-\omega)\nonumber\\
&+&\Sigma^{(a)}_{2}(-{\bf k},-\omega)g({\bf k},\omega)],
\end{eqnarray}
respectively, with the mean-field (MF) dressed charge carrier
diagonal Green's function \cite{ma} $g^{(0)}({\bf k},\omega)=
\omega-\xi_{\bf k}$, where the MF dressed charge carrier excitation
spectrum $\xi_{\bf k}=Zt\chi_{1}\gamma_{\bf k}-Zt'
\chi_{2}\gamma_{\bf k}'-\mu$, with $\gamma_{\bf k}=(1/Z)
\sum_{\hat{\eta}}e^{i{\bf k}\cdot{\hat{\eta}}}$, ${\gamma_{\bf k}
}'=(1/Z)\sum_{\hat{\tau}}e^{i{\bf k}\cdot{\hat{\tau}}}$, the spin
correlation function $\chi_{2}=\langle S^{+}_{i}
S^{-}_{i+\hat{\tau}}\rangle$, Z is the number of the nearest
neighbor or second-nearest neighbor sites, while the dressed charge
carrier self-energy functions are given by \cite{feng,ma},
\begin{eqnarray}
\Sigma^{(a)}_{1}(&{\bf k}&,i\omega_n)={1\over N^2}\sum_{{\bf p},
{\bf p'}}(Zt\gamma_{{\bf p}+{\bf p}'+{\bf k}}-Zt'\gamma_{{\bf p}
+{\bf p}'+{\bf k}})^{2}\nonumber\\
&\times&{1\over\beta}\sum_{ip_m}g({\bf p} +{\bf k}
,ip_{m}+i\omega_{n})\Pi({\bf p'},{\bf p},ip_{m}), \\
\Sigma^{(a)}_{2}(&{\bf k}&,i\omega_n)={1\over N^2}\sum_{{\bf p},
{\bf p}'}(Zt\gamma_{{\bf p}+{\bf p}'+{\bf k}}-Zt'\gamma_{{\bf p}
+{\bf p}'+{\bf k}})^{2}\nonumber\\
&\times&{1\over\beta}\sum_{ip_m}\Im({\bf p}+{\bf k}
,ip_{m}+i\omega_{n})\Pi({\bf p'},{\bf p},ip_{m}),
\end{eqnarray}

with the spin pair bubble $\Pi({\bf p'},{\bf p},ip_{m})=(1/\beta)
\sum_{ip'_{m}}D^{(0)}({\bf p'},ip'_{m})D^{(0)}({\bf p'+p},ip'_{m}
+ip_{m})$, $N$ is the number of sites, and the MF spin Green's
function \cite{ma},
\begin{eqnarray}
D^{(0)}({\bf p},\omega)={B_{\bf p}\over 2\omega_{\bf p}}\left({1
\over\omega-\omega_{\bf p}}-{1\over\omega+\omega_{\bf p}}\right),
\end{eqnarray}
where $B_{{\bf p}}=2\lambda_{1}(A_{1}\gamma_{{\bf p}}-A_{2})-
\lambda_{2}(2\chi^{z}_{2}\gamma_{{\bf p }}'-\chi_{2})$,
$\lambda_{1}=2ZJ_{{\rm eff}}$, $\lambda_{2}=4Z\phi_{2}t'$, $A_{1}=
\epsilon\chi^{z}_{1}+\chi_{1}/2$, $A_{2}=\chi^{z}_{1}+\epsilon
\chi_{1}/2$, $\epsilon=1+2t\phi_{1}/J_{{\rm eff}}$, the dressed
charge carrier's particle-hole parameters $\phi_{1}=\langle
a^{\dagger}_{i\sigma}a_{i+\hat{\eta}\sigma}\rangle$ and $\phi_{2}=
\langle a^{\dagger}_{i\sigma}a_{i+\hat{\tau}\sigma}\rangle$, the
spin correlation functions $\chi^{z}_{1}=\langle S_{i}^{z}
S_{i+\hat{\eta}}^{z}\rangle$ and $\chi^{z}_{2}=\langle S_{i}^{z}
S_{i+\hat{\tau}}^{z}\rangle$, and the MF spin excitation spectrum,
\begin{eqnarray}
\omega^{2}_{{\bf p}}&=& \lambda_{1}^{2}[(A_{4}-\alpha\epsilon
\chi^{z}_{1}\gamma_{{\bf p}}-{1\over 2Z}\alpha\epsilon\chi_{1})
(1-\epsilon\gamma_{{\bf p}})\nonumber\\
&+&{1\over 2}\epsilon(A_{3}-{1\over 2} \alpha\chi^{z}_{1}-\alpha
\chi_{1}\gamma_{{\bf p}})(\epsilon-
\gamma_{{\bf p}})] \nonumber \\
&+&\lambda_{2}^{2}[\alpha(\chi^{z}_{2}\gamma_{{\bf p}}'-{3\over
2Z}\chi_{2})\gamma_{{\bf p}}'+{1\over 2}(A_{5}-{1\over 2}\alpha
\chi^{z}_{2})]\nonumber\\
&+& \lambda_{1}\lambda_{2}[\alpha\chi^{z}_{1}(1-\epsilon
\gamma_{{\bf p}})\gamma_{{\bf p}}'+{1\over 2}\alpha(\chi_{1}
\gamma_{{\bf p}}'-C_{3})(\epsilon- \gamma_{{\bf p}}) \nonumber\\
&+&\alpha \gamma_{{\bf p}}'(C^{z}_{3}-\epsilon \chi^{z}_{2}
\gamma_{{\bf p}})-{1\over 2}\alpha\epsilon(C_{3}- \chi_{2}
\gamma_{{\bf p}})],
\end{eqnarray}
with $A_{3}=\alpha C_{1}+(1-\alpha)/(2Z)$, $A_{4}=\alpha C^{z}_{1}
+(1-\alpha)/(4Z)$, $A_{5}=\alpha C_{2}+(1-\alpha)/(2Z)$, and the
spin correlation functions
$C_{1}=(1/Z^{2})\sum_{\hat{\eta},\hat{\eta'}}\langle
S_{i+\hat{\eta}}^{+}S_{i+\hat{\eta'}}^{-}\rangle$,
$C^{z}_{1}=(1/Z^{2})\sum_{\hat{\eta},\hat{\eta'}}\langle
S_{i+\hat{\eta}}^{z}S_{i+\hat{\eta'}}^{z}\rangle$,
$C_{2}=(1/Z^{2})\\
\sum_{\hat{\tau},\hat{\tau'}}\langle
S_{i+\hat{\tau}}^{+}S_{i+\hat{\tau'}}^{-}\rangle$,
$C_{3}=(1/Z)\sum_{\hat{\tau}}\langle S_{i+\hat{\eta}}^{+}
S_{i+\hat{\tau}}^{-}\rangle$, and $C^{z}_{3}=(1/Z)
\sum_{\hat{\tau}}\langle S_{i+\hat{\eta}}^{z}
S_{i+\hat{\tau}}^{z}\rangle$. In order to satisfy the sum rule of
the correlation function $\langle S^{+}_{i}S^{-}_{i}\rangle=1/2$ in
the case without AFLRO, an important decoupling parameter $\alpha$
has been introduced in the MF calculation \cite{feng,ma}, which can
be regarded as the vertex correction \cite{kondo}.

In the framework of the kinetic energy driven superconductivity
\cite{feng}, the self-energy function $\Sigma^{(a)}_{2}({\bf k},
\omega)$ describes the effective dressed charge carrier pair gap
function, while the self-energy function $\Sigma^{(a)}_{1} ({\bf
k},\omega)$ renormalizes the MF dressed charge carrier spectrum,
and therefore it describes the quasiparticle coherence. As in the
hole-doped case \cite{guo}, we only discuss the low-energy
behavior of the electron-doped cuprate superconductors, therefore
the effective dressed charge carrier pair gap function and
quasiparticle coherent weight can be discussed in the static
limit. In this case, we follow the previous discussions for the
hole-doped case \cite{guo}, and obtain explicitly the dressed
charge carrier diagonal and off-diagonal Green's functions of the
electron-doped cuprate superconductors as,
\begin{eqnarray}
g({\bf k,\omega})&=&Z^{(a)}_{F}{U_{a{\bf k}}^2\over\omega- E_{a{\bf
k}}}+Z^{(a)}_{F}{V_{a{\bf k}}^2\over\omega+E_{a{\bf k}}},\\
\Im^{+}({\bf k},\omega)&=&-Z^{(a)}_{F}{\bar{\Delta}_{aZ}({\bf k})
\over 2E_{a{\bf k}}}\left({1\over\omega-E_{a{\bf k}}}\right.
\nonumber\\
&-&\left. {1\over\omega+E_{a{\bf k}}}\right),
\end{eqnarray}
where the dressed charge carrier quasiparticle coherence factors
$U_{a{\bf k}}^{2}=(1+\bar\xi_{\bf k}/E_{a{\bf k}})/2$ and $V_{a{\bf
k}}^{2}=(1-\bar\xi_{\bf k}/E_{a{\bf k}})/2$, the dressed charge
carrier quasiparticle coherent weight $Z^{(a)-1}_{F}=1-
\Sigma^{(a)}_{1o}({\bf k},\omega=0)\mid_{{\bf k}=[\pi,0]}$, the
renormalized dressed charge carrier excitation spectrum $\bar
\xi_{\bf k}=Z^{(a)}_{F}(\xi_{\bf k}+\Sigma^{(a)}_{1e})$ with
$\Sigma^{(a)}_{1e} = \Sigma^{(a)}_{1e}({\bf k},\omega=0) \mid_{{\bf
k }=[\pi,0]}$, the renormalized dressed charge carrier pair gap
function $\bar{\Delta}_{aZ}({\bf k})=Z^{(a)}_{F}
\bar{\Delta}_{a}({\bf k})$ with $\bar{\Delta}_{a}({\bf k})=
\Sigma^{(a)}_{2}({\bf k},\omega=0)$, and the dressed charge carrier
quasiparticle spectrum $E_{a{\bf k}} = \sqrt{\bar{\xi}_{\bf
k}^{2}+\mid\bar{\Delta}_{aZ}({\bf k}) \mid^{2}}$, while
$\Sigma^{(a)}_{1e}({\bf k},\omega)$ and $\Sigma^{(a)}_{1o} ({\bf
k},\omega)$ are the corresponding symmetric and antisymmetric parts
of the self-energy function $\Sigma^{(a)}_{1}({\bf k},\omega)$. As
we have mentioned above, the electron-doped cuprate superconductors
are characterized by an overall d-wave pairing symmetry
\cite{shen2,sato}. Based on the kinetic energy driven SC mechanism
\cite{feng}, we \cite{ma} have shown within the $t$-$t'$-$J$ model
that the electron Cooper pairs of the electron-doped cuprate
superconductors have a dominated d-wave symmetry. In this case, we
consider the d-wave case of the electron-doped cuprate
superconductors, i.e., $\bar{\Delta}_{a} ({\bf
k})=\bar{\Delta}_{a}\gamma^{(d)}_{{\bf k}}$, with
$\gamma^{(d)}_{{\bf k}}=({\rm cos}k_{x}-{\rm cos}k_{y})/2$. With the
help of the above discussions, the dressed charge carrier effective
gap parameter and quasiparticle coherent weight in Eqs. (7) and (8)
satisfy the following two equations,
\begin{eqnarray}
1&=&{1\over N^3}\sum_{{\bf k},{\bf q},{\bf p}}(Zt\gamma_{{\bf k}
+{\bf q}}-Zt'\gamma_{{\bf k}+{\bf q}}')^{2}\gamma_{{\bf k}-{\bf q}
+{\bf p}}^{(d)}\gamma_{\bf k}^{(d)}\nonumber\\
&\times&{Z^{(a)2}_{F}B_{\bf q}B_{\bf p} \over E_{a\bf k}\omega_{\bf
q}\omega_{\bf p}}\left ({F_{1}^{(1)} ({\bf k},{\bf q},{\bf p})\over
(\omega_{\bf p}-\omega_{\bf q})^{2} -E^{2}_{a\bf k}}\right.\nonumber\\
&-&\left.{F_{1}^{(2)}({\bf k},{\bf q},{\bf p})\over(\omega_{\bf p}
+\omega_{\bf q})^{2}-E^{2}_{a\bf k}}\right ), \\
{1\over Z^{(a)}_{F}}&=&1+{1\over N^{2}}\sum_{{\bf q},{\bf p}}(Zt
\gamma_{{\bf p}+{\bf k}_0}-Zt'\gamma_{{\bf p}+{\bf k}_0}')^{2}
Z^{(a)}_{F}\nonumber\\
&\times&{B_{\bf q}B_{\bf p}\over 4\omega_{\bf q}\omega_{\bf
p}}\left({F_{2}^{(1)}({\bf q},{\bf p})\over (\omega_{\bf p}-
\omega_{\bf q}-E_{a{\bf p}-{\bf q}+{\bf k}_0})^2}\right.\nonumber\\
&+&\left. {F_{2}^{(2)} ({\bf q},{\bf p})\over (\omega_{\bf
p}-\omega_{\bf q}+E_{a{\bf p}-
{\bf q}+ {\bf k}_0})^{2}}\right.\nonumber \\
&+& \left .{F_{2}^{(3)}({\bf q},{\bf p})\over (\omega_{\bf p}
+\omega_{\bf q}-E_{a{\bf p}-{\bf q}+{\bf k}_0})^{2}}\right.
\nonumber\\
& +& \left . {F_{2}^{(4)}({\bf q},{\bf p})\over (\omega_{\bf p}+
\omega_{\bf q}+E_{a{\bf p}-{\bf q}+{\bf k}_0})^{2}}\right),
\end{eqnarray}
respectively, where ${{\bf k}_0}=[\pi,0]$, $F_1^{(1)}({\bf k}, {\bf
q},{\bf p})=(\omega_{{\bf p}}-\omega_{{\bf q}})[n_{B} (\omega_{{\bf
q}})-n_{B}(\omega_{{\bf p}})][1-2n_{F}(E_{a{\bf k}}) ]+E_{a{\bf k}}[
n_{{\bf B}}(\omega_{{\bf p}})n_{{\bf B}} (-\omega_{{\bf q}})+
n_{{\bf B}}(\omega_{{\bf q}})n_{{\bf B}} (-\omega_{{\bf p}})]$,
$F_1^{(2)}({\bf k,q,p})\\
=(\omega_{{\bf p}} +\omega_{{\bf q}})[n_{B}(-\omega_{{\bf
p}})-n_{B}(\omega_{{\bf q} })][1-2n_{F} (E_{a{\bf k}})]+E_{a{\bf
k}}[ n_{{\bf B}} (\omega_{{\bf p}})n_{{\bf B}}(\omega_{{\bf
q}})+n_{{\bf B}} (-\omega_{{\bf p}})n_{{\bf B}} (-\omega_{{\bf
q}})]$, $F_2^{(1)}({\bf q},{\bf p})=n_{F}(E_{{a{\bf p}-{\bf q}+{\bf
k}_0}} )[n_{B}(\omega_{{\bf q}})-n_{B}(\omega_{{\bf p}})]-n_{{\bf
B}} (\omega_{{\bf p}})n_{{\bf B}}(-\omega_{{\bf q}})$, $F_2^{(2)}
({\bf q},{\bf p})=n_{F}(E_{{a\bf p-q+k_0}}) [n_{B}(\omega_{{\bf p}}
)-n_{B}(\omega_{{\bf q}})]-n_{{\bf B}}(\omega_{{\bf q}})n_{{\bf B}
}(-\omega_{{\bf p}})$, $F_2^{(3)}({\bf q},{\bf p})= n_{F}(E_{{a{\bf
p}-{\bf q}+{\bf k}_0}})[n_{B}(\omega_{{\bf q}})- n_{B}(-\omega_{{\bf
p}})]+n_{{\bf B}}(\omega_{{\bf p}})n_{{\bf B}} (\omega_{{\bf q}})$,
$F_2^{(4)}({\bf q},{\bf p})=n_{F}(E_{{a{\bf p }-{\bf q}+{\bf k}_0}})\\
\times[n_{B}(-\omega_{{\bf q}})-n_{B}(\omega_{{\bf p}})]+n_{{\bf
B}}(-\omega_{{\bf p}})n_{{\bf B}}(-\omega_{{\bf q}}) $, and
$n_{B}(\omega)$ and $n_{F}(\omega)$ are the boson and fermion
distributions, respectively. These two equations must be solved
simultaneously with other self-consistent equations \cite{feng,ma},
then all order parameters, decoupling parameter $\alpha$, and
chemical potential $\mu$ are determined by the self-consistent
calculation.

For the understanding of the electronic state properties of the
electron-doped cuprates in the SC-state, we need to calculate the
electron diagonal and off-diagonal Green's functions $G(i-j,t-t')=
\langle\langle C_{i\sigma}(t);C^{\dagger}_{j\sigma}(t')\rangle
\rangle$ and $\Gamma^{\dagger} (i-j,t-t')=\langle\langle
C^{\dagger}_{i\uparrow}(t);C^{\dagger}_{j\downarrow}(t')\rangle
\rangle$, which are the convolutions of the spin Green's function
and dressed charge carrier diagonal and off-diagonal Green's
functions in the CSS fermion-spin theory, and reflect the
charge-spin recombination \cite{anderson}. According to the MF spin
Green's function (9) and dressed charge carrier diagonal and
off-diagonal Green's functions (11) and (12), we can obtain the
electron diagonal and off-diagonal Green's functions as,
\begin{eqnarray}
G({\bf k},\omega)&=&{1\over N}\sum_{{\bf p}}Z_{F}{B_{{\bf p}}\over
2\omega_{{\bf p}}}\left\{{\rm coth}[{\beta\omega_{{\bf p}}\over 2}]
\right.\nonumber \\
&\times& \left. \left ( {U^{2}_{a{\bf p}+{\bf k}}\over\omega-
E_{a{\bf p} +{\bf k}}-\omega_{\bf p}}+{U^{2}_{a{\bf p}+{\bf k}}
\over\omega-E_{a{\bf p}+{\bf k}}+\omega_{\bf p}} \right. \right.
\nonumber\\
&+&\left.\left.{V^{2}_{a{\bf p}+{\bf k}}\over\omega+E_{a{\bf p}
+{\bf k}}-\omega_{\bf p}}+{V^{2}_{a{\bf p}+{\bf k}}\over \omega+
E_{a{\bf p}+{\bf k}}+\omega_{\bf p}}\right)\right.\nonumber\\
&+&\left.{\rm tanh}[{\beta E_{a{\bf p}+{\bf k}}\over 2}]
\left({U^{2}_{a{\bf p}+{\bf k}}\over \omega-E_{a{\bf p}+{\bf k}}
-\omega_{\bf p}} \right.\right.\nonumber\\
&-&\left.\left.{U^{2}_{a{\bf p}+{\bf k}}\over\omega-E_{a{\bf p}+
{\bf k}}+\omega_{\bf p}}+{V^{2}_{a{\bf p}+{\bf k}}\over\omega+
E_{a{\bf p}+{\bf k}}+\omega_{\bf p}}\right.\right. \nonumber\\
&-&\left.\left.{V^{2}_{a{\bf p}+{\bf k}}\over\omega +E_{a{\bf p}
+{\bf k}}-\omega_{\bf p}}\right)\right\},\\
\Gamma^{\dag}({\bf k},\omega)&=&{1\over N}\sum_{{\bf p}}Z_{F}
{\bar\Delta_{aZ}({\bf p}+{\bf k})\over 2E_{a{\bf p}+{\bf k}}}
{B_{{\bf p}}\over 2\omega_{{\bf p}}}\left\{{\rm coth}
[{\beta\omega_{{\bf p}}\over 2}]\right.\nonumber\\
&\times& \left. \left({1\over\omega-E_{a{\bf p}+ {\bf k}}-
\omega_{{\bf p}}} + {1\over \omega-E_{a{\bf p}+{\bf k}}+
\omega_{{\bf p}}}\right.\right. \nonumber\\
&-&\left.\left.{1\over\omega+E_{a{\bf p}+{\bf k}}+ \omega_{{\bf p}}}
-{1\over\omega+E_{a{\bf p}+{\bf k}}-\omega_{{\bf p}}} \right)
\right.\nonumber\\
&+&\left.{\rm tanh}[{\beta E_{a{\bf p}+{\bf k}}\over 2})
\left({1\over\omega-E_{a{\bf p}+{\bf k}}-\omega_{{\bf p}}}
\right.\right.\nonumber\\
&-&\left.\left.{1\over\omega-E_{a{\bf p}+{\bf k}}+\omega_{{\bf p}}
}-{1\over\omega+E_{a{\bf p}+{\bf k}}+\omega_{{\bf p}}}\right.
\right.\nonumber\\
&+& \left. \left. {1\over \omega + E_{a{\bf p}+{\bf k}}-
\omega_{{\bf p}}}\right)\right\},
\end{eqnarray}
respectively, with the electron quasiparticle coherent weight
$Z_{F}=Z^{(a)}_{F}/2$,  then the electron spectral function $A({\bf
k},\omega)=-2{\rm Im}G({\bf k},\omega)$ and SC gap function
$\Delta({\bf k})=-(1/\beta)\sum_{i\omega_n}\Gamma^{\dag} ({\bf
k},i\omega_n)$ are obtained as,
\begin{eqnarray}
A({\bf k},\omega)&=&2\pi{1\over N}\sum_{{\bf p}}Z_{F}{B_{{\bf p}}
\over 2\omega_{{\bf p}}}\{{\rm coth}[{\beta\omega_{{\bf p}}\over 2
}]\nonumber\\
&\times& [U^{2}_{a{\bf p} + {\bf k}}\delta(\omega - E_{a{\bf p}+
{\bf k}} -\omega_{\bf p})\nonumber\\
&+&U^2_{a{\bf p+k}}\delta(\omega-E_{a{\bf p}+{\bf k}}
+\omega_{\bf p})\nonumber\\
&+& V^2_{a{\bf p}+{\bf k}}\delta(\omega+E_{a{\bf p}+{\bf k}}-
\omega_{\bf p})\nonumber\\
&+&V^{2}_{a{\bf p}+{\bf k}}\delta(\omega+E_{a{\bf p}+
{\bf k}}+\omega_{\bf p})]\nonumber \\
&+&{\rm tanh}[{\beta E_{a{\bf p}+{\bf k}}\over 2}][U^{2}_{a{\bf p}
+{\bf k}}\delta(\omega-E_{a{\bf p}+{\bf k}}-\omega_{\bf p})
\nonumber\\
&-& U^2_{a{\bf p}+{\bf k}}\delta(\omega-E_{a{\bf p}+{\bf k}}+
\omega_{\bf p}) \nonumber\\
&+&V^2_{a{\bf p}+{\bf k}}\delta(\omega+E_{a{\bf p}+{\bf k}}+
\omega_{\bf p})\nonumber\\
&-&V^2_{a{\bf p}+{\bf k}}\delta(\omega+E_{a{\bf p}+
{\bf k}}-\omega_{\bf p})]\},\\
\Delta({\bf k})&=&-{1\over N}\sum_{{\bf p}}Z_{F}{\bar\Delta_{Za}
({\bf p}-{\bf k})\over E_{a{\bf p}-{\bf k}}}{\rm tanh}[{\beta
E_{a{\bf p}-{\bf k}}\over 2}]\nonumber\\
&\times&{B_{{\bf p}}\over 2\omega_{{\bf p}}} {\rm coth}
[{\beta\omega_{{\bf p}}\over 2}],
\end{eqnarray}
respectively. From Eq. (18), the SC gap parameter in Eq. (4) can be
evaluated as $\Delta=-\chi_{1}\Delta_{a}$. As in the hole-doped case
\cite{feng}, both dressed charge carrier (then electron) pairing gap
parameter and pairing interaction in the electron-dopede cuprate
superconductors are doping dependent. In this case, the experimental
observed doping dependence of the SC gap parameter should be an
effective SC gap parameter $\bar{\Delta}=-\chi_{1}\bar{\Delta}_{a}$.
For a complement of the previous analysis of superconductivity in
the electron-doped cuprate superconductors \cite{ma}, we plot (a)
the effective SC gap parameter $\bar{\Delta}$ at temperature
$T=0.002J$ and (b) the SC transition temperature $T_{c}$ as a
function of the doping concentration for parameters $t/J=-2.5$ and
$t'/t=0.3$ in Fig. 1. For comparison, the corresponding experimental
results of the SC gap parameter \cite{qazilbash} and SC transition
temperature \cite{peng} of the electron-doped cuprate
superconductors as a function of the doping concentration are also
shown in Fig. 1(a) and 1(b), respectively. Our present results
indicate that in analogy to the phase diagram of the hole-doped
case, superconductivity appears over a narrow range of doping in the
electron-doped cuprate superconductors. As shown in the
self-consistent equations in Eqs. (13) and (14), the SC-state of the
electron-doped cuprate superconductors is controlled by both SC gap
function and quasiparticle coherence \cite{feng,ma}, which leads to
that the SC transition temperature increases with increasing doping
in the underdoped regime, and reaches a maximum in the optimal
doping, then decreases sharply with increasing doping in the
overdoped regime. However, the maximum achievable SC transition
temperature in the optimal doping in the electron-doped cuprate
superconductors is much lower than that of the hole-doped case due
to the electron-hole asymmetry. Although we focus on the
quasiparticle coherent weight at the antinodal point in the above
discussions, our present results of the doping dependence of the
effective SC gap parameter and SC transition temperature are
consistent with these of the previous results \cite{ma}, where it
has been focused on the quasiparticle coherent weight near the nodal
point.

\begin{figure}
\begin{center}
\includegraphics*[height=8cm,width=8cm]{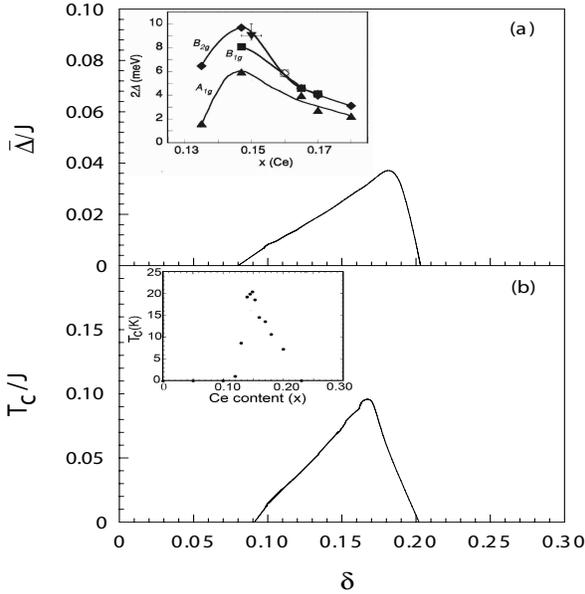}
\end{center}
\caption{(a) The effective SC gap parameter $\bar{\Delta}$ at
$T=0.002J$ and (b) the SC transition temperature $T_{c}$ as a
function of the doping concentration for $t/J=-2.5$ and $t'/t=0.3$.
Inset: the corresponding experimental results of the electron doped
cuprate superconductors taken from Refs. [21] and [4].}
\end{figure}

Now we turn to discuss the electron structure of the electron-doped
cuprate superconductors. We have performed a calculation for the
electron spectral function (17), and the results of $A({\bf
k},\omega)$ in the $[\pi,0]$ point with the doping concentration
$\delta=0.11$ (solid line), $\delta=0.13$ (dashed line), and
$\delta=0.15$ (dotted line) at $T=0.002J$ for $t/J=-2.5$ and
$t'/t=0.3$ are plotted in Fig. 2 in comparison with the
corresponding experimental result \cite{shen3} of the electron-doped
cuprate superconductor Nd$_{1.85}$Ce$_{0.15}$CuO$_{4}$ (inset). From
Fig. 2, we therefore find that there is a sharp SC quasiparticle
peak near the electron Fermi energy in the $[\pi,0]$ point. In
particular, this electron spectrum is doping dependent. In analogy
to the hole-doped case \cite{guo}, the spectral weight of the SC
quasiparticle peak in the electron-doped cuprate superconductors
increases with increasing the doping concentration, while the
position of the SC quasiparticle peak is shifted towards the Fermi
energy. Furthermore, we have discussed the temperature dependence of
the electron spectrum, and the results show that the spectral weight
of the SC quasiparticle peak decreases as temperature is increased.
Our these results are in qualitative agreement with the experimental
data \cite{shen,shen2,shen3,shen4}.

\begin{figure}
\begin{center}
\includegraphics*[height=6cm,width=7.5cm]{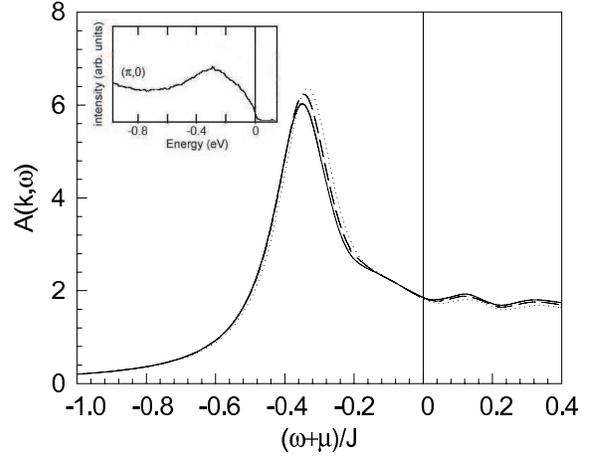}
\end{center}
\caption{The electron spectral function $A({\bf k},\omega)$ in the
$[\pi,0]$ point with $\delta=0.11$ (solid line), $\delta=0.13$
(dashed line), and $\delta=0.15$ (dotted line) at $T=0.002J$ for
$t/J=-2.5$ and $t'/t=0.3$. Inset: the corresponding experimental
result of the electron-doped cuprate superconductor
Nd$_{1.85}$Ce$_{0.15}$CuO$_{4}$ taken from Ref. [8].}
\end{figure}

\begin{figure}
\begin{center}
\includegraphics*[height=6cm,width=7.5cm]{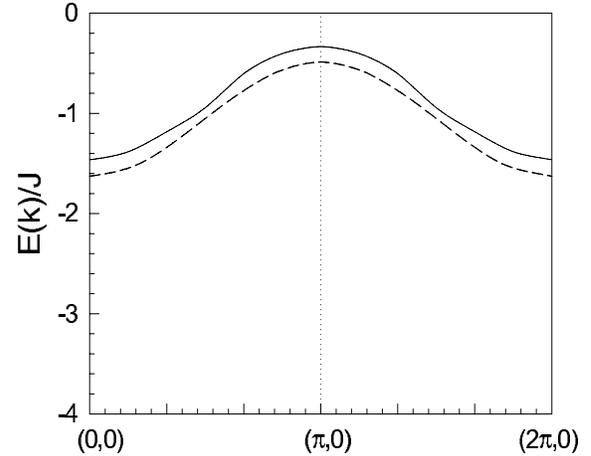}
\end{center}
\caption{The positions of the lowest energy SC quasiparticle peaks
in $A({\bf k},\omega)$ as a function of momentum along the direction
$[0,0]\rightarrow [\pi,0]\rightarrow [2\pi,0]$ with $\delta=0.15$ at
$T=0.002J$ for $t/J=-2.5$ and $t'/t=0.3$. The dashed line is
corresponding result [12] of the electron dispersion of the
hole-doped cuprate superconductors at $\delta=0.15$ with $T=0.002J$
for parameters $t/J=2.5$ and $t'/t=0.3$.}
\end{figure}

For the further understanding physical properties of the SC
quasiparticles near the $[\pi,0]$ point, we have made a series of
calculations for $A({\bf k},\omega)$ around the $[\pi,0]$ point, and
the results show that as in the hole-doped case, the sharp SC
quasiparticle peak in the electron-doped cuprate superconductors
persists in a very large momentum space region around the $[\pi,0]$
point. To show this point clearly, we plot the positions of the
lowest energy SC quasiparticle peaks in $A({\bf k},\omega)$ as a
function of momentum along the direction $[0,0] \rightarrow
[\pi,0]\rightarrow [2\pi,0]$ at $\delta=0.15$ with $T=0.002J$ for
$t/J=-2.5$ and $t'/J=0.3$ in Fig. 3 (solid line). For comparison,
the corresponding result \cite{guo} of the lowest energy SC
quasiparticle peaks in the electron spectral function of the
hole-doped cuprate superconductors at $\delta=0.15$ with $T=0.002J$
for $t/J=2.5$ and $t'/t=0.3$ is also plotted in Fig. 3 (dashed
line). From Fig. 3, it is shown that the sharp SC quasiparticle
peaks around the $[\pi,0]$ point at low energies disperse very
weakly with momentum, which also is corresponding to the unusual
flat band appeared in the normal-state around the $[\pi,0]$ point
\cite{shen6,guo1}. In comparison with the hole doped case
\cite{guo}, our results also show explicitly that although the
electron-hole asymmetry is observed in the phase diagram
\cite{shen,shen1}, the electronic structure of the electron-doped
cuprates in the SC state is similar to that in the hole-doped case.

Within the framework of the kinetic energy driven d-wave
superconductivity, the essential physics of the electronic structure
in the electron-doped cuprate superconductors is the same as that in
the hole-doped case \cite{guo}. The SC-state of the electron-doped
cuprate superconductors is the conventional
Bardeen-Cooper-Schrieffer (BCS) like \cite{bcs}, and the SC
quasiparticle has the Bogoliubov-quasiparticle nature. This can be
understood from the electron diagonal and off-diagonal Green's
functions in Eqs. (15) and (16). As in the hole-doped case
\cite{guo}, the spins center around the $[\pi,\pi]$ point in the MF
level \cite{ma}, therefore the main contributions for the spins
comes from the $[\pi,\pi]$ point, where $\omega_{{\bf p}=[\pi,
\pi]}\sim 0$. In this case, the electron diagonal and off-diagonal
Green's functions in Eqs. (15) and (16) can be approximately reduced
in terms of the self-consistent equation \cite{ma} $1/2= \langle
S_{i}^{+}S_{i}^{-}\rangle=(1/N)\sum_{{\bf p}}B_{{\bf p}} {\rm
coth}(\beta \omega_{{\bf p}}/2)/(2\omega_{{\bf p}})$ as,
\begin{eqnarray}
G({\bf k},\omega) &\approx& Z_{F}{U_{{\bf k}}^{2}\over\omega-
E_{{\bf k}}}+Z_{F}{V_{{\bf k}}^{2}\over \omega+E_{{\bf k}}},\\
\Gamma^{\dag}({\bf k},\omega) &\approx& Z_{F}{\bar{\Delta}_{aZ}
({\bf k})\over 2E_{{\bf k}}}\left({1\over \omega-E_{{\bf k}}}
-{1\over\omega+E_{{\bf k}}}\right),
\end{eqnarray}
where the electron quasiparticle coherence factors $U^{2}_{{\bf k}
}\approx V^{2}_{a{\bf k+k_{A}}}$ and $V^{2}_{{\bf k}}\approx
U^{2}_{a{\bf k+k_{A}}}$, the electron quasiparticle spectrum
$E_{{\bf k}}\approx E_{a{\bf k+k_{A}}}$, and ${\bf k_{A}}=
[\pi,\pi]$, which show that the dressed charge carrier quasiparticle
coherence factors $V_{a{\bf k}}$ and $U_{a{\bf k}}$ and
quasiparticle spectrum $E_{a{\bf k}}$ have been transferred into the
electron quasiparticle coherence factors $U_{{\bf k}}$ and $V_{{\bf
k}}$ and quasiparticle spectrum $E_{{\bf k}}$, respectively, by the
convolutions of the spin Green's function and dressed charge carrier
diagonal and off-diagonal Green's functions due to the charge-spin
recombination. This also reflects that in the kinetic energy driven
SC mechanism, the dressed charge carrier pairs condense with the
d-wave symmetry, then the electron Cooper pairs originating from the
dressed charge carrier pairing state are due to the charge-spin
recombination, and their condensation automatically gives the
electron quasiparticle character. This is why the basic BCS
formalism \cite{bcs} is still valid in discussions of the doping
dependence of the effective SC gap parameter and SC transition
temperature, and the SC coherence of the quasiparticle peak in the
electron-doped cuprate superconductors, although the pairing
mechanism is driven by the kinetic energy by exchanging spin
excitations, and other exotic magnetic scattering \cite{dai} is
beyond the BCS theory. On the other hand, although there is a
similar strength of the magnetic interaction $J$ for both hole-doped
and electron-doped cuprates, the interplay of $t'$ with $t$ and $J$
causes a further weakening of the AF spin correlation for the hole
doping, and enhancing the AF spin correlation for the electron
doping \cite{gooding}, which shows that the AF spin correlations in
the electron doping is stronger than these in the hole-doped side.
This may lead to the charge carrier's localization over a broader
range of doping for the electron doping. As a consequence, the
asymmetry of the electron spectrum in the hole-doped and
electron-doped cuprates emerges. This is also why superconductivity
appears over a narrow range of doping in the electron-doped
cuprates. In the normal-state, we \cite{guo1} have shown within the
CSS fermion-spin theory that the lowest energy states are located at
the $[\pi,0]$ point for the electron doping, this means that at low
doping, the Fermi surface is an electron-pocket centered at the
$[\pi,0]$ point, and then further electron doping may lead to the
creation of a new holelike Fermi surface centered at the $[\pi,\pi]$
point, in qualitative agreement with the experimental data
\cite{shen4}. As we have mentioned above, the most contributions of
the electronic states in the SC-state for the electron doping come
from $[\pi,0]$ point \cite{shen,shen1,shen2,shen3,shen4}, and then
superconductivity is characterized by an overall d-wave pairing
symmetry $\bar{\Delta}({\bf k})=\bar{\Delta}({\rm cos}k_{x}-{\rm
cos} k_{y})/2$ \cite{shen2,sato}. In this case, the d-wave SC gap,
and therefore the electron pairing energy scale, is maximized at
$[\pi,0]$ point.

In summary, we have discussed the electronic structure of the
electron-doped cuprate superconductors based on the kinetic energy
driven d-wave SC superconductivity. Our results show explicitly
that although the electron-hole asymmetry is observed in the phase
diagram, the electronic structure of the electron-doped cuprates
in the SC state is similar to that in the hole-doped case. With
increasing the electron doping, the spectral weight in the
$[\pi,0]$ point increases, while the position of the sharp SC
quasiparticle peak is shifted towards the Fermi energy. In analogy
to the hole-doped case, the SC quasiparticles around the $[\pi,0]$
point disperse very weakly with momentum.

\begin{ack}
This work was supported by the National Natural Science Foundation
of China under Grant No. 90403005, and the funds from the Ministry
of Science and Technology of China under Grant No. 2006CB601002.
\end{ack}

\end{document}